# An accumulator model for primes and targets with independent response activation rates: Basic equations for average response times

## arXiv.com/1804.08513 [q-bio.NC]

Thomas Schmidt
Experimental Psychology Unit
University of Kaiserslautern
Kaiserslautern, Germany

Filipp Schmidt
Department of General and Experimental Psychology
University of Giessen
Giessen, Germany

**Abstract**

**In response priming tasks, speeded responses are performed toward target stimuli preceded by prime stimuli. Responses are slower and error rates are higher when prime and target are assigned to different responses, compared to assignment to the same response, and those priming effects increase with prime-target SOA. Here, we generalize Vorberg et al.'s (2003) accumulator model of response priming, where response activation is first controlled exclusively by the prime and then taken over by the actual target. Priming thus occurs by motor conflict because a response-inconsistent prime can temporarily drive the process towards the incorrect response. While the original model assumed prime and target signals to be identical in strength, we allow different rates of response activation (cf. Mattler & Palmer, 2012; Schubert et al., 2012). Our model predicts that stronger primes mainly increase priming effects in response times and error rates, whereas stronger targets mainly diminish response times and priming effects.**

*1. Introduction*
*Response priming* (Klotz & Neumann, 1999; Klotz & Wolff, 1995; Vorberg, Mattler, Heinecke, Schmidt, & Schwarzbach, 2003) is a response-conflict paradigm that serves to study the time-course of sequential response activation (Schmidt et al., 2011). On each trial, observers respond to a target stimulus by pressing a left or right key (e.g., to indicate whether the target is red or green; Schmidt, 2000). The target is preceded by a prime stimulus which is either mapped to the same response as the target (*consistent prime*; e.g., red target preceded by red prime) or to the alternative response (*inconsistent prime*; e.g., red target preceded by green prime). Typically, consistent primes will speed responses to the target and produce fewer errors while inconsistent primes will slow responses and produce more errors. This difference between response times or error rates in consistent versus inconsistent trials is labeled the *priming effect*.

The magnitude of this priming effect critically depends on the stimulus-onset asynchrony (SOA), that is, the time between the onsets of prime and target. Generally, response priming occurs for SOAs up to about 100 ms and increases approximately linearly with SOA (Vorberg et al., 2003). Accumulator models of response priming (Vorberg et al., 2003; cf. Mattler & Palmer, 2012; Schubert, Palazova, & Hutt, 2013) assume that the prime starts activating the response assigned to it, followed by response activation by the target after the SOA. Depending on whether the prime is consistent or inconsistent with the target, it will drive the response process into either the correct or incorrect direction, and may even cause a response error. The longer the SOA, the more time is available for the prime to activate the response, and the larger the priming effect in both response times and error rates (e.g., Vorberg et al., 2003; Klotz, Heumann, Ansorge, & Neumann, 2007; Leuthold & Kopp, 1998; Schmidt, 2002; Vath & Schmidt, 2007).

Furthermore, the magnitude of response priming depends on the strength of the prime and the target in activating the response. Prime and target strength can be controlled by luminance or color contrast (Vath & Schmidt, 2007), familiarity (Kiefer, Sim, & Wentura, 2015; Pfister, Pohl, Kiesel, & Kunde, 2012), stimulus eccentricity (Lingnau & Vorberg, 2005), similarity of stimulus alternatives (Schmidt & Schmidt, 2013, 2014), by spatial attention to the prime (Kiefer & Brendel, 2006; Schmidt & Seydell, 2008), or by feature-based attention to the prime (Schmidt & Schmidt, 2010; Tapia, Breitmeyer, & Shooner, 2010). Reversely, the prime can be weakened by degradation of its response-defining features (Schmidt, Weber, & Schmidt, 2014). Generally, stronger primes lead to larger priming effects, and stronger targets lead to faster overall responses but smaller priming effects (Schmidt, Haberkamp, & Schmidt, 2011).

In the following, we will describe our generalization of Vorberg et al.'s (2003) accumulator model of response priming, and describe the basic behavior of the model.

### 2. The free-rate accumulator model

We define two counters, $n_1(t)$ and $n_2(t)$, gathering evidence for leftward or rightward responses to a prime followed by a target (Fig. 1). Counters are defined such that $n_1(t)$ gathers information for the response afforded by the prime, and $n_2(t)$ for the opposite response. Therefore, when the prime affords a leftward response, $n_1(t)$ is the leftward counter and $n_2(t)$ is the rightward counter. To follow the accumulation process through time, we distinguish two time periods. Let $t = 0$ mark the time when the accumulation process for prime-related information starts, and let $s$ be the prime-target SOA. Then, for $0 \leq t < s$ the process is entirely controlled by the prime. When the target arrives at time $s$, it takes over the accumulation process from the state that it was in at time $t = s$.

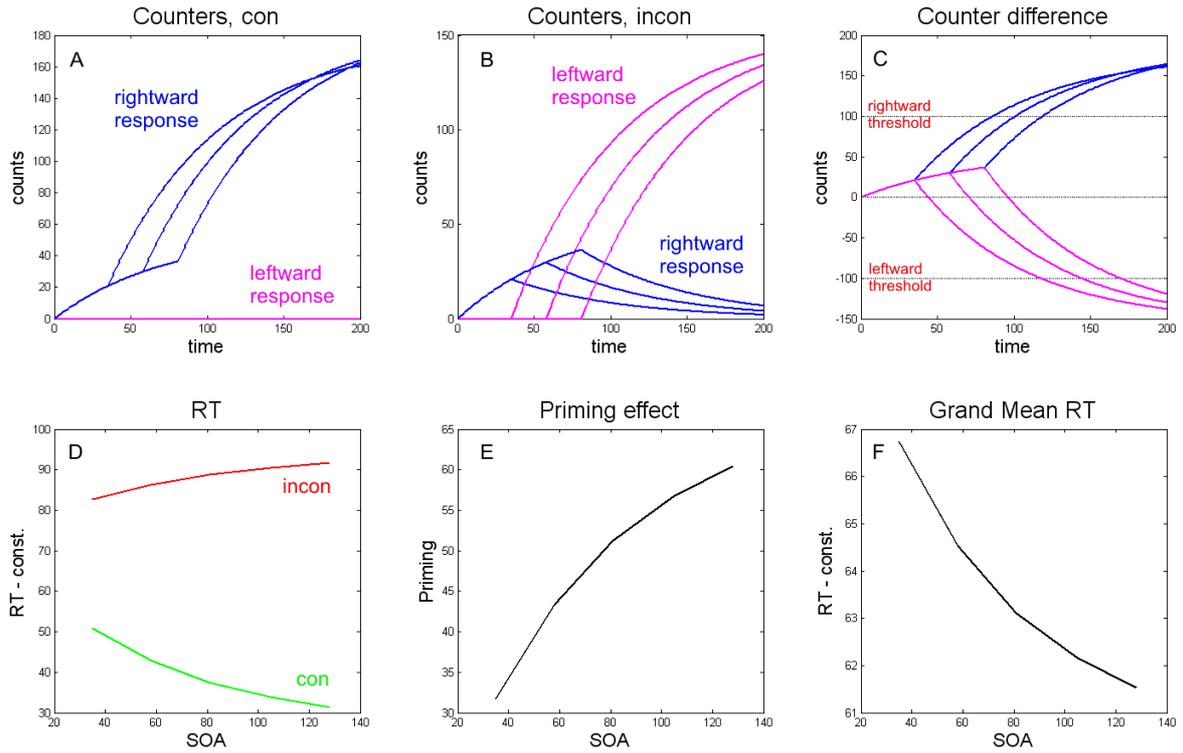

Fig. 1A: In consistent trials, counter $n_1$ (blue) starts gathering evidence for the response afforded by the prime. When the target arrives at time $s$, it proceeds in the same direction at a new rate. Counter $n_2$ (pink) remains inactive because there is no input to which it is tuned. Three different SOAs are shown. B: In inconsistent trials, the initial activation of $n_1$ decays after time $s$ because there is no further input, while $n_2$ starts getting active after delay $s$. C: Responses are registered when the difference between the counters crosses the threshold $\pm c$. D: Response time effects as a continuous function of SOA, now relative to target onset. We have to subtract a constant containing all the slack due to conduction times, nonmotor processing time, mechanical inertia etc. E: Priming function, $P(s)$. F: Response-speed function, $R(s)$, also defined up to the slack constant.

A counter is incremented by a stimulus to which it is tuned; it is decremented by random decay. Accumulation and decay occur in discrete counts, and each increment or decrement is a Poisson event. The resulting behavior is an *immigration-death process*, where the count at time $t$ is distributed as a Poisson variable with mean and variance $n_i(t)$. Therefore, when a counter is activated by relevant information, the expectancy of the accumulated evidence for its respective response is

$$n_i(t) = \frac{1}{\nu} \lambda (1 - e^{-\nu t}) \quad,$$

where $\lambda$ (read: lambda) is the rate at which evidence accumulates and $\nu$ (read: nu) is the rate at which it decays. If the counter is no longer activated by any relevant information after time $s$, its current activation level $n_i(s)$ decays passively to zero according to

$$n_i(t-s) = n_i(s) e^{-\nu(t-s)} \quad.$$

(Read $t - s$ as "time since target started to accumulate".) In fact, this is all the mathematical arsenal we need, and everything else follows from these basic equations. The logic of the model is the following (Vorberg et al., 2003). Each counter gathers evidence for its respective response over time. The time when the *difference* $n_1(t) - n_2(t)$ between the counters crosses a threshold $c$, the response afforded by the prime is given; when it crosses the threshold $-c$, the opposite response is given.

In a consistent trial, the prime will first activate $n_1$ at rate $\lambda_P$ until time $t = s$, the point in time when the prime is replaced by the target in the accumulation process. Then, starting from time $t = s$, $n_1$ is further activated, but now at rate $\lambda_T$. Because $n_2$ is not activated by any stimulus, it remains at resting level, $n_2(t) = 0$. Therefore, for $t < s$:

$$n_1(t) = \frac{1}{\nu} \lambda_P (1 - e^{-\nu t}) \ ;$$

$$n_2(t) = 0 \ ;$$

and for $t \geq s$:

$$n_1(t) = n_1(s) + \frac{1}{\nu} \lambda_T (1 - e^{-\nu(t-s)})$$
$$= \frac{1}{\nu} \lambda_P (1 - e^{-\nu s}) + \frac{1}{\nu} \lambda_T (1 - e^{-\nu(t-s)})$$
$$= \frac{1}{\nu} [\lambda_P (1 - e^{-\nu s}) + \lambda_T (1 - e^{-\nu(t-s)})] \ ;$$

$$n_2(t) = 0 \ .$$

The response is given as soon as the difference $n_1(t) - n_2(t)$ crosses the threshold $c$. To find this crossing time, we just have to set set $n_1(t)$ equal to $c$ and solve for time. Assuming the response time is longer than the prime-target SOA, this time is:

$$t \geq s:$$
$$d_{con}(t') = n_1(t') - n_2(t') = c$$

$$\frac{1}{\nu} [\lambda_P (1 - e^{-\nu s}) + \lambda_T (1 - e^{-\nu(t'-s)})] - 0 = c$$

$$\Rightarrow t' = t'_{con} = \frac{1}{\nu} \log \frac{\lambda_T e^{\nu s}}{\lambda_P (1 - e^{-\nu s}) + \lambda_T - \nu c} \ , \quad \text{(Eq. 1a)}$$

where the log symbol denotes the natural logarithm, and the prime in $t'$ indicates that we are now looking at a particular value of the variable $t$.

The inconsistent case is more involved. In an inconsistent trial, the prime will again activate $n_1$ at rate $\lambda_P$ until time $t = s$. However, because the prime is inconsistent with the target, it activates the wrong counter. When the target arrives at time $t = s$, it starts activating the correct counter, $n_2$, at rate $\lambda_T$, while activation in $n_1$ starts to decay at rate $\nu$. Therefore, for $t < s$:

$$n_1(t) = \frac{1}{\nu}\lambda_P(1-e^{-\nu t})$$

$$n_2(t) = 0 \; ;$$

and for $t \geq s$:

$$n_1(t) = n_1(s)e^{-\nu(t-s)} = \frac{1}{\nu}\lambda_P(1-e^{-\nu s})e^{-\nu(t-s)}$$

$$= \frac{1}{\nu}\lambda_P(e^{\nu s}-1)e^{-\nu t} \; ;$$

$$n_2(t) = \frac{1}{\nu}\lambda_T(1-e^{-\nu(t-s)}) \; .$$

Because the difference $n_1(t) - n_2(t)$ starts into the wrong direction, when the inconsistent target takes over the process, it has to start from state $n_1(s)$, which is way positive, and then move the process all the way down towards the *opposite* threshold, $-c$. It is this time loss that generates the priming effect. Again assuming that the response time is longer than the prime-target SOA:

$t \geq s$:

$$d_{incon}(t') = n_1(t') - n_2(t') = -c$$

$$\frac{1}{\nu}\lambda_P(e^{\nu s}-1)e^{-\nu t'} - \frac{1}{\nu}\lambda_T(1-e^{-\nu(t'-s)}) = -c$$

$$\frac{1}{\nu}\{[(\lambda_P+\lambda_T)e^{\nu s} - \lambda_P]e^{-\nu t} - \lambda_T\} = -c$$

$$\Rightarrow t' = t'_{incon} = \frac{1}{\nu}\log\frac{(\lambda_P+\lambda_T)e^{\nu s}-\lambda_P}{\lambda_T - \nu c} \; . \qquad (\text{Eq. 1b})$$

Note that $t'_{con}$ and $t'_{incon}$ are defined relative to the onset of the counting process triggered by the prime. To define them relative to target onset, as is common in response time research, we have to subtract the SOA, $s$. Further note that response times as measured in the experiment involve some additional processing time due to conduction delays, mechanical slack, and other factors, which we assume to be constant. Therefore, empirical response times $t_{con}$ and $t_{incon}$ are related to the model values by $t_i = t'_i - s + const.$

If we define response time as a function of SOA in consistent and inconsistent trials as $C(s)$ and $I(s)$, respectively, we can decompose the entire response time pattern into two effects. First, the *priming function* is defined as

$$P(s) = I(s) - C(s) = t'_{incon}(s) - t'_{con}(s)$$

$$= \frac{1}{\nu} \log \frac{[(\lambda_P+\lambda_T)e^{\nu s}-\lambda_P][\lambda_P(1-e^{-\nu s})+\lambda_T-\nu c]}{\lambda_T(\lambda_T-\nu c)e^{\nu s}} \quad .$$  (Eq. 2a)

Second, the *response speed function* is defined as

$$R(s)=\frac{1}{2}[I(s)+C(s)]=\frac{1}{2}[t'_{incon}(s)+t_{con}(s)]$$

$$=\frac{1}{2\nu}\log\frac{\lambda_T(\lambda_P+\lambda_T)e^{2\nu s}-\lambda_P\lambda_T e^{\nu s}}{(\lambda_T-\nu c)\lambda_P(1-e^{-\nu s})+(\lambda_T-\nu c)^2} \quad .$$  (Eq. 2b)

The decomposition can be helpful because $\lambda_T$ and $\lambda_P$ have separable effects on response speed and priming. Note that both functions depend on all four parameters. The functions are related by:

$$I(s)=R(s)+\frac{1}{2}P(s) \quad ;$$

$$C(s)=R(s)-\frac{1}{2}P(s) \quad .$$

Note that priming effects are symmetric about *R(s)* simply because *R(s)* is defined as the average of *I(s)* and *C(s)*. This does not imply that inconsistent and consistent primes induce symmetric costs and benefits in absolute time, which is generally not the case. Also remember that, just as *t'_con* and *t'_incon*, *R(s)* is only defined up to a constant. This assumption of constant residual time may be violated by several factors, like foreperiod or attentional effects.

### 3. Basic behavior of the model

The four free parameters of the model all have separable effects on the data.

- $\lambda_P$ *(evidence accumulation by the prime)* mainly controls the magnitude of priming effects in both response times and error rates. The larger $\lambda_P$, the deeper the motor conflict induced by inconsistent primes, and the larger the priming effect in response times and error rates.
- $\lambda_T$ *(evidence accumulation by the target)* both controls the magnitude of priming effects and the overall speed of responses. The larger $\lambda_T$, the faster the overall response and the smaller the priming effect (because stronger targets are quicker in counteracting the effect of an inconsistent prime).
- $\nu$ *(decay rate)* controls the curvature of all functions. If response times for consistent and inconsistent trials depend linearly on SOA, so that the curvature is low, then $\nu$ can be assumed to be small relative to $\lambda_P$ and $\lambda_T$. Larger $\nu$ leads to negatively accelerating functions for response times and priming, and it can also increases overall response times.
- *c (response threshold)* affects the overall response time and the error rate. It is intuitively clear that a lower response threshold allows inconsistent primes to induce more errors, and at shorter SOAs.

     7

Unfortunately, it is quite difficult to model error rates in closed form, and we did not attempt it here. Instead, we are currently running simulations of the underlying Poisson process to simulate not only error rates but also response time distributions. It turns out that if $c$ is set to a high value (e.g., 100 counts, as is customary in the literature), the stochastic process is so fine-grained that errors at short SOAs become very unlikely. To have a smooth, linear increase in error rate with SOA, a very low response threshold is needed (10 counts or even less).

It is important to consider the special case of simultaneous presentation of primes and targets, because this is the standard setting in some response conflict paradigms closely related to response priming (like the Eriksen and the Stroop effect). In our model, setting $s$ to 0 yields identical predictions for consistent and inconsistent response times that do no longer depend on prime strength:

$$t'_{con}(s=0) = t'_{incon}(s=0) = \frac{1}{\nu} \log \left( \frac{\lambda_T}{(\lambda_T - \nu c)} \right) ,$$

so that $P(s = 0) = 0$. (Setting $s = 0$ in Eq. 3a gives the same result, $P(s = 0) = [\log (\lambda_T / \lambda_T)]/\nu = 0$.) Simultaneous presentation of primes and targets (e.g., arranged as flankers) is thus predicted to generate no response priming, contrary to empirical findings from flanker tasks. In general, temporally overlapping primes and targets may require explicit modeling of joint response activation, such as a prime and target that together activate a response at rate $\lambda_T + \lambda_P$ (consistent trials) or $\lambda_T - \lambda_P$ (inconsistent trials) for the time period in question.

## 4. Recommendations for model fitting

The model has four free parameters: $\lambda_P$, $\lambda_T$, $\nu$, and $c$. The full power of the model can be harnessed in a parametric research design where priming functions are traced through variations in prime or target strength. If only a single priming function has to be predicted, the full four-parameter model will probably overfit the data. For that purpose, we recommend restricting the parameter space. First, the response threshold $c$ can be set to a low, constant value (e.g., 10 counts). Second, if the response-time and priming functions have little curvature, $\nu$ can be set to a small constant value as well (e.g., 0.001). For this range of values, plausible initial values for $\lambda_P$ and $\lambda_T$ are between 0 and 1.

Keep in mind that the parameters have to obey the condition that $\lambda_T > \nu c$. Otherwise, the priming function is undefined either because of division by zero or because of taking the log of a negative value. This side condition makes sense because $\nu c$ is an upper limit of how many counts can potentially decay (namely, if the process is very close to the response threshold). If $\lambda_T$ is smaller than that value, then the rate of decay exceeds the rate of accumulation, and the threshold cannot be reached.

Consistent and inconsistent response times should always be estimated jointly, not separately. To do this, observe that $t'_{incon} = t'_{con} + P$, where $P$ is the priming effect. Let $D$ be a dummy variable that equals 1 for inconsistent trials and 0 otherwise. Then, functions can be fitted to $t'_{con} + DP(s)$ in a single step. A similar trick can be used to test the difference of accumulation between two conditions, for example, two different primes or targets. Let *D* be a dummy variable that equals 1 for condition 1 and 0 otherwise. Then, replacing $\lambda$ with $\lambda + D(\lambda + \delta)$ enables us to estimate $\delta$, the difference in processing rates.

Remember that model fitting and model testing should be separated. One straightforward technique to cross-validate the fit is to estimate parameters with one half of the data set (e.g., odd-numbered trials) and then fit it to the other half to do the statistical test. An even more elegant technique is bootstrapping. Generally, model parameters should be estimated separately for each participant.

*5. Relation to previous models*

Our model is based on the general principle of *direct parameter specification* (Neumann, 1990), which states that single stimulus parameters can trigger preprogrammed choice responses without the need for awareness. More specifically, it is an instance of our *chase theory of response priming* (Schmidt et al., 2011), which states that prime and target activate their response as a sequence of neuronal feedforward sweeps (Lamme & Roelfsema, 2000) independent of visual awareness of the prime (Schmidt & Vorberg, 2006; Schmidt, 2007). Please keep in mind that our model only describes the response conflict that occurs for prime-target SOAs up to about 100 ms. For longer SOAs, additional processes come into play, especially processes of response inhibition (Panis & Schmidt, 2016; Schmidt, Hauch, & Schmidt, 2015). Response inhibition can lead to reversal of the priming effect, with faster responses in inconsistent than in consistent trials (*negative compatibility effect*; Eimer & Schlaghecken, 1998; Lingnau & Vorberg, 2005).

The response-time model we describe here is a straightforward generalization of Vorberg et al.'s (2003) accumulator model to arbitrary, independent processing rates of prime and target. Vorberg et al. (2003) assume that prime and target are processed by the same rate, $\lambda = \lambda_P = \lambda_T$, and that the system does not distinguish between primes and targets: When a consistent target appears, it does not refresh the accumulation process, but the process simply continues at the same rate of deceleration as if the prime had never changed. This simplification leads to an extremely economical model: The priming function depends only on $v$ but is independent of $c$ and $\lambda$. However, since the decay parameter $v$ is a system constant that does not depend on the stimulus, the model cannot explain how priming functions depend on stimulus parameters.

Mattler and Palmer (2012) published a simulation model that makes the same generalization as ours but assumes that response thresholds become narrower during the trial. Our explicit model of the response times should basically correspond to their simulations, but there may be details in their simulation that do not exactly fit our modeling here. Schubert, Palazova, and Hutt (2013) have presented a model that makes the exact same generalization we attempt here, but additionally proposes that the prime processing rate depends on the prime-target interval. To facilitate comparisons between those models, we use the same mathematical notation as previous papers.